\documentclass[
aps,%
12pt,%
final,%
notitlepage,%
oneside,%
onecolumn,%
nobibnotes,%
nofootinbib,%
superscriptaddress,%
noshowpacs]%
{revtex4}
\usepackage{graphicx}
\usepackage{color}

\newcommand{\be}{\begin{eqnarray}}
\newcommand{\ee}{\end{eqnarray}}

\begin{document}

\noindent {\it Astronomy Reports, 2021, Vol. 98, No. 1}
\bigskip\bigskip  \hrule\smallskip\hrule
\vspace{35mm}


\title{PRIMORDIAL BLACK HOLES AND MODIFICATION OF ZELDOVICH-NOVIKOV MECHANISM
\footnote{Paper presented at the Fourth Zeldovich 
meeting, an international conference in honor of Ya. B. Zeldovich held in Minsk, Belarus on September 7--11, 2020. Published by the 
recommendation of the special editors: S. Ya. Kilin, R. Ruffini and G. V. Vereshchagin.}}

\author{\bf \copyright $\:$  2021.
\quad \firstname{A.~D.}~\surname{Dolgov}}%
\email{dolgov@fe.infn.it}
\affiliation{Novosibirsk State University, Novosibirsk, Russia}%

\author{\bf \firstname{K.~A.}~\surname{Postnov}}
\email{kpostnov@gmail.com}
\affiliation{Sternberg Astrnomikical Institute, Moscow, Russia}


\begin{abstract}

\centerline{\footnotesize Received: ;$\;$
Revised: ;$\;$ Accepted: .}\bigskip\bigskip\bigskip


A review of the recent astronomical observations is presented and it is argued that the data strongly indicate 
that practically all observed black holes are primordial. A modified mechanism of the primordial black hole
 formation is described. The log-normal mass spectrum predicted by this mechanism is strongly confirmed
 by the LIGO data on gravitational wave registration.
\end{abstract}

\maketitle

\section{Introduction}

The idea of the primordial black hole (PBH)  i.e. of black holes which could be 
created in  the early universe during pre-stellar epoch 
was first put forward by Ya. Zeldovich and I. Novikov in 1996 in the seminal paper entitled
{ "The Hypothesis of Cores Retarded During Expansion and the Hot Cosmological Model"~\cite{ZN-BH}.
{ According to their idea, the 
density contrast in the early universe inside the bubble with the radius equal to the cosmological horizon 
might accidentally happen to be large, {${\delta \rho /\rho \approx 1}$,} then
that piece of volume would be inside its gravitational radius i.e. it became  a PBH, which
decoupled  from the cosmological expansion. }
This idea was  elaborated later by S.~Hawking in his paper "Gravitationally collapsed objects of very low 
mass"~\cite{SH-BH} and by  B. Carr and S. Hawking in "Black holes in the early Universe''~\cite{CH-BH}.

The mass inside horizon in the early universe at the radiation dominated stage (RD) stage, ${r_{hor} = 2 t}$ is equal to
 $M_{hor} = m_{Pl}^2 t $ and if $ \delta \rho /\rho = 1$, then ${M_{BH} =  M_{hor}}$ and the gravitational radius is
indeed equal to horizon, $ r_g = {2M}/{m_{Pl}^2} = r_{hor}$.

In ref.~\cite{AD-JS}  the mechanism of PBH formation was modified in several essential features. It was the first suggestion
to employ cosmological  inflation for PBH formation. It allowed to create PBHs with masses exceeding millions solar masses.
According to the proposed mechanism large isocurvature density perturbations had been generated during inflationary stage
which later on, at the QCD phase transition, turned into density perturbations. The mechanism of ref.~\cite{AD-JS}, see also
\cite{DKK}, predicted very simple log-normal mass spectrum of PBH:
\be
\frac{dN}{dM} = \mu^2 \exp{[-\gamma \ln^2 (M/M_0)]}.
\label{dn-dM}
\ee
Parameters $\mu$ and $\gamma$ are determined by physics at very high energies and unknown, but  $M_0$ should be
 equal to the mass inside horizon at the QCD phase transition~\cite{AD-KP-mass}:
 \be
 M_{hor}  = 8 M_\odot \cdot  \left(\frac{100\,{\rm MeV}}{T}\right)^2,
\ee
where $M_\odot = 2\cdot 10^{33} $ g is the solar mass.
According to the lattice calculations $T_{QCD} = 100 -150 $ MeV but with large chemical potential of quarks 
$T_{QCD}$ may be noticeably 
smaller, so we expect $M_0$ somewhat larger than $10 M_\odot$, which happens to be in very good agreement with
observations, see figures below.

Subsequent mechanism of inflationary creation of PBH was studied about two years later in ref.~\cite{INN}. It is based on
suggestion of ref.~\cite{aas-pert} on the generation of the  perturbation of the inflaton field in a model with double field inflation.
The PBH mass spectrum calculated in ref.~\cite{INN} has rather  complicated  analytical structure and to the best of my
knowledge was not applied to the analysis of the observational data.

At the present time there is a large lot of inflationary mechanisms of PBH creation with variety of mass spectra.

\section{Modified mechanism of PBH creation \label{s-PBH-mdf}}

According to  suggestion of ref.~\cite{AD-JS} the condition of PBH formation was prepared during inflation
by dynamical generation of large baryonic number perturbations at  astrophysical scales. 
It was achieved by the popular model of Affleck and Dine (AD)~\cite{Aff-Dine} 
baryogenesis which allows to create 
baryon asymmetry of order of unity, much larger than the observed conventional one, 
${{ \beta \approx 6\cdot 10^{-10}}}$.

Due to interaction between  the inflaton and the scalar Affleck-Dine field with nonzero baryon number the  creation of 
domains with large $\beta$ could take place only during relatively short time and thus
cosmologically small but possibly astronomically large 
bubbles with high ${ \beta}$ could be created.
By construction they occupy {a small
fraction of the universe volume,} while the rest of the universe has normal
${{ \beta \approx 6\cdot 10^{-10}}}$. 

Thus during inflation huge isocurvature perturbations in baryonic number were arranged which
transformed into density perturbations at the
QCD phase transition when massless quarks turned into heavy baryons. 
 Inflationary prehistory allows for  
creation of  huge PBH with masses up to  ${(10^4-10^5) M_\odot}$ and possibly even higher.

Note that a possible outcome of this mechanism is the creation of early compact star-like objects which
might consist both of matter and antimatter.

\section{Log-normal spectrum, comparison with observations}

Very powerful way to gain information on the mass spectrum of PBHs is opened by the LIGO/Virgo data on gravitational
wave emission by black hole binaries.

As is well known, a binary system of  two rotating gravitationally bound massive bodies should  emit gravitational 
waves. In quasi-stationary inspiral regime, the radius of the orbit and the rotation frequency
slowly change and the GW frequency is approximately equal to the double Newtonian rotation frequency:
\be 
\omega^2_{orb} =  \frac{M_1+M_2}{m_{Pl}^2 R^3}\,.
\label{omega-GE}
\ee
{The luminosity of the GW radiation during this stage is equal to:}
\be 
L  = \frac{32}{5}\,m_{Pl}^2\left(\frac{M_c\,\omega_{orb}}{m_{Pl}^2}\right)^{10/3}\,,
\label{GW-lum}
\ee
where $M_1$, $M_2$ are the masses of two rotating bodies  and 
${M_c}$ is the so called chirp mass: 
\be 
M =\frac{(M_1\,M_2)^{3/5}}{(M_1+M_2)^{1/5}} \, ,
\label{M-chirp}
\ee

In ref.~\cite{chirp} the integrated chirp mass distribution is calculated on the basis of the publicly available LIGO data:
\be
F_{PBH} (<M) = \int_0^M P_{PBH} (M) dM = \frac{\int_0^M dM \cal{ D R(M)} }{\int_0^\infty dM \cal{ D R(M)} },
\label{F-chirp}
\ee
where $\cal{ D R(M)}$ is the detection rate per year. Note that this ratio does not depend upon the normalization factor
$\mu$ of distribution (\ref{dn-dM}). The comparison of the LIGO data with theoretical expectation for $F_{PBH} (<M)$ according
to ref.~\cite{chirp} for the best fit of spectrum parameters $M_0$ and $\gamma$ is presented in Fig.~\ref{fig-1}.

\begin{figure}[!h]
\centering
\includegraphics[width=.48\textwidth]{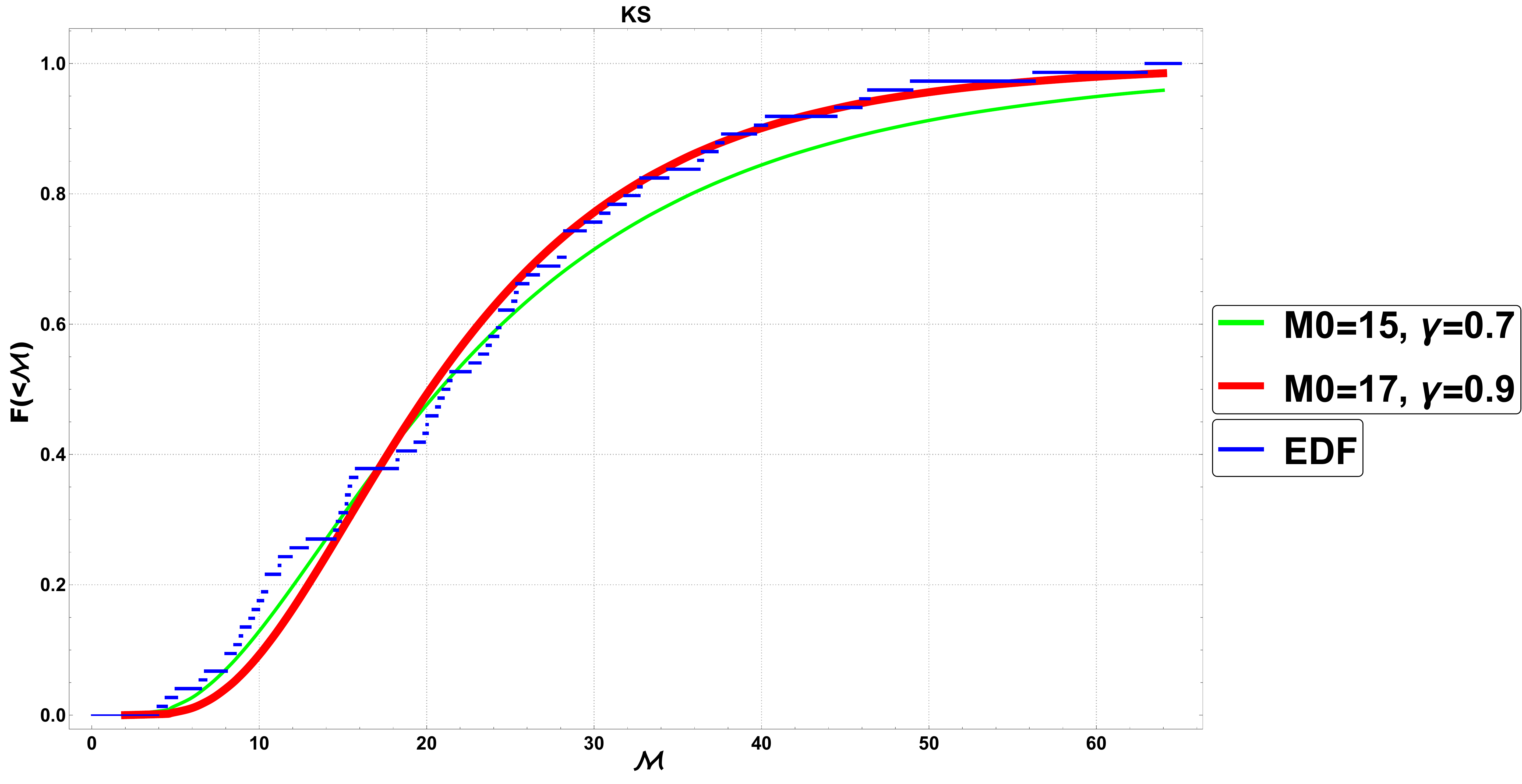}
\includegraphics[width=.48\textwidth]{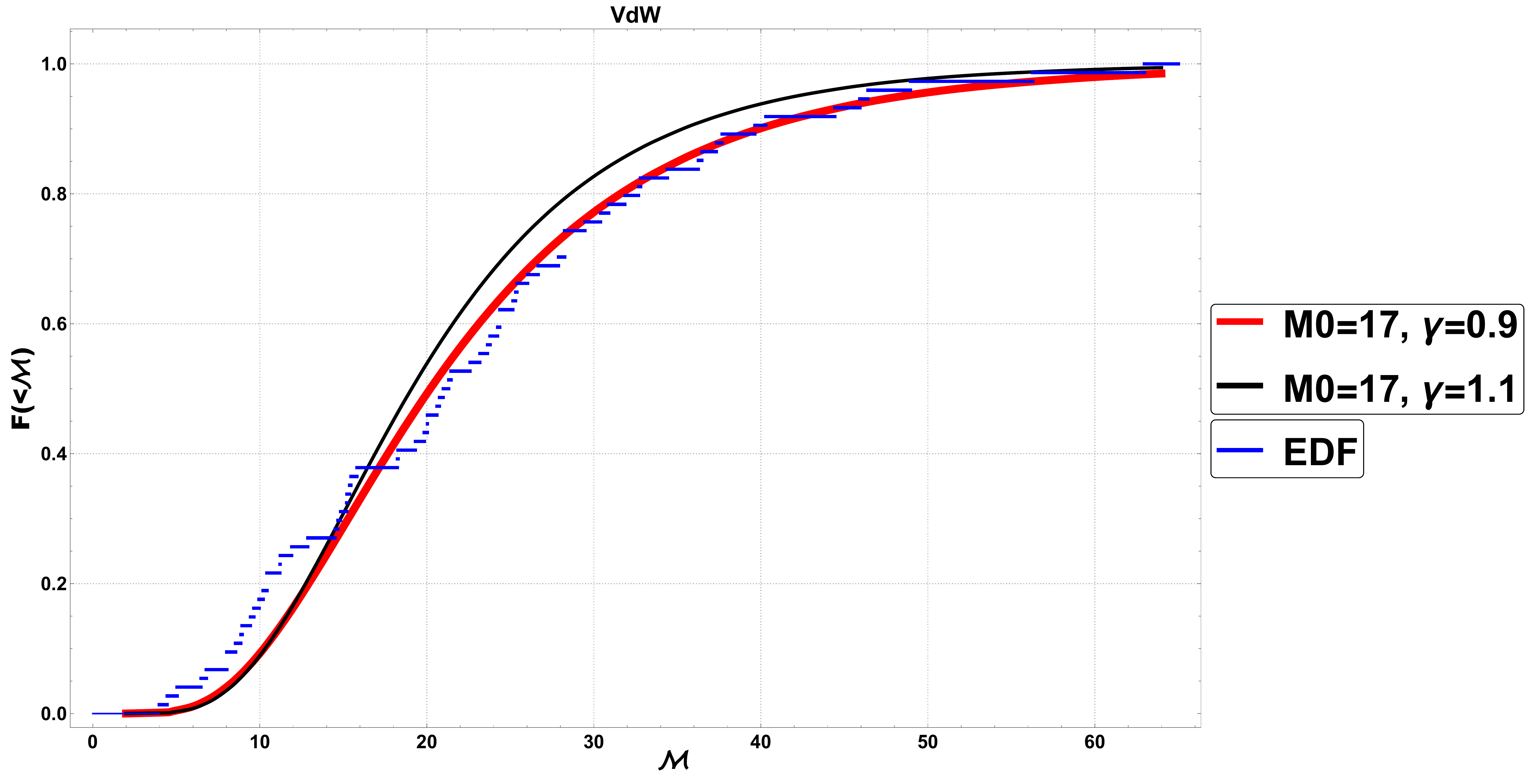} 
\caption{Empirical distribution function (EDF) comparison to the log-normal spectrum with the best fit 
 parameters $M_0$ and $\gamma$; 
left panel: model distribution $F_{PBH}(< M)$ for two best 
Kolmogorov-Smirnov tests;  right panel: model distribution $F_{PBH}(< M)$  for two best Van der Waerden tests.
}
\label{fig-1}
\end{figure}

\begin{figure}[!h]
\begin{center}
\includegraphics[scale=0.2]{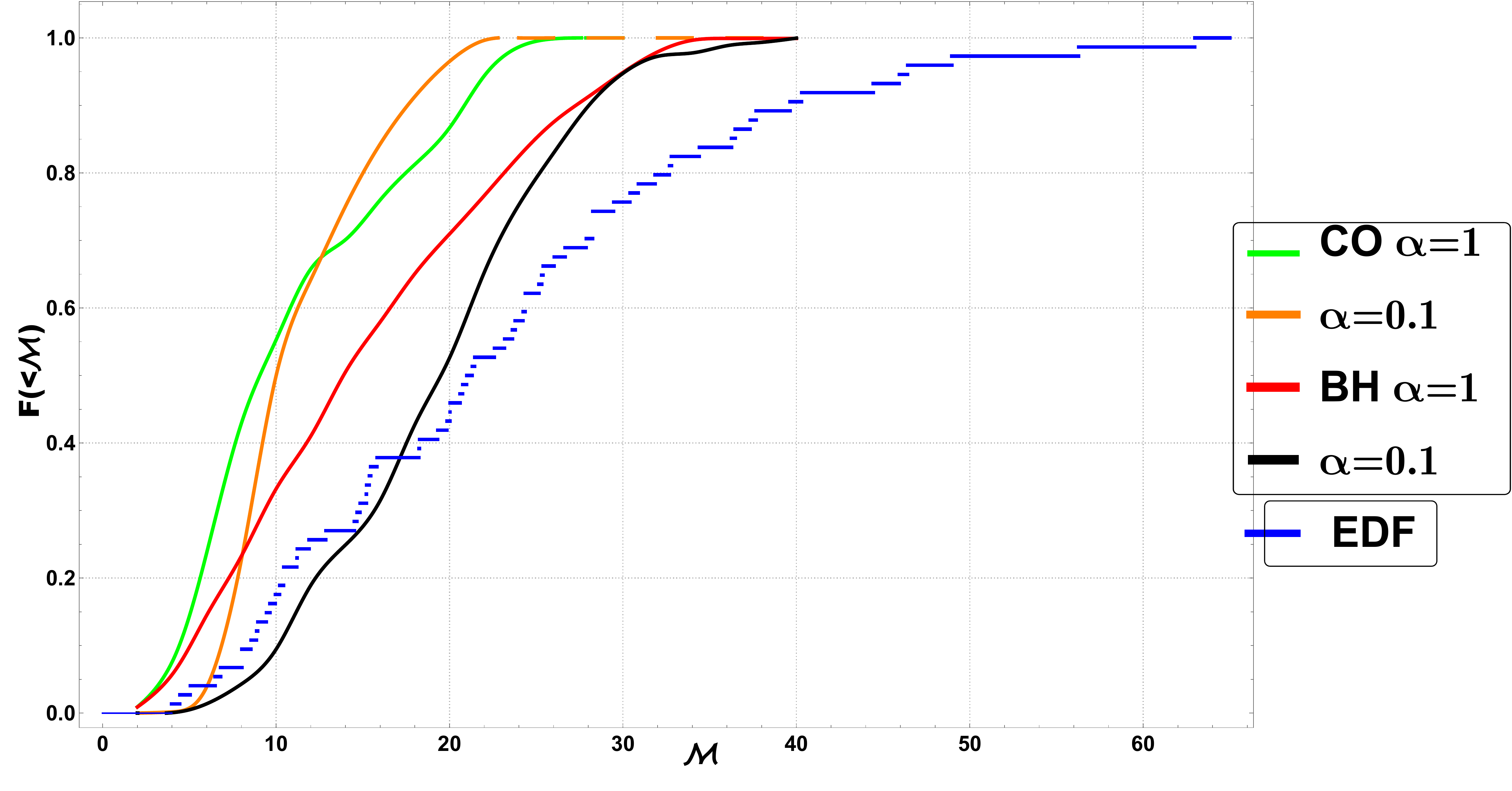}
\caption{Cumulative distributions $F(< M)$ for several { astrophysical} models of binary BH coalescences, see ref.~\cite{chirp}. }
\end{center}
\label{fig-2}
\end{figure}


According to these figures the log-normal mass spectrum with the central mass $M_0 = 17 M_\odot$ perfectly agrees with the data.
On the other hand, similar comparison with the best fit of several astrophysical models of BH origin, shown in Fig. 2,
demonstrates quite poor description.

An analysis of the recently released LIGO data leads to the same value of $M_0$ and to somewhat smaller $\gamma$, i.e. 
$\gamma =0.7$ instead of  $\gamma \approx 1$.

To conclude this section, the available data present strong evidence in favor of primordial origin of BH binaries observed by LIGO
through registration of gravitational waves from their coalescence. Log-normal mass spectrum of PBHs is clearly confirmed by observations.
To the best of the author knowledge, no other spectrum is checked by the data, to say nothing of  confirmation. 
The puzzle of the hundred solar mass black hole  finds natural solution if one assumes that it is primordial and no exotic mechanism of its formation is necessary.  The discovery of such "impossible" BH is an impressive confirmation of our model.

More arguments in favor of primordial formation of LIGO sources can be found e.g. in ref.~\cite{BDPP}.

With the determined here parameters of the log-normal mass spectrum of PBHs (see also \cite{AD-SP})
one may conclude that universe should be full of PBHs 
in all mass ranges from supermassive with masses up to 100 billion solar masses, down to intermediate mass BHs, with
$M = (10^2 -10^5) M_\odot$, and of course plenty of BHs with stellar masses.
It would not be an exaggeration to say that (almost) all BHs in the universe are primordial.  These black holes may naturally
solve multiple cosmological problems accumulated in the present day and early universe~\cite{AD-UFN,Carr-conundra},
as is discussed in the following section.

\section{Cosmic conundra and PBH}

Primordial black holes which for many years have been taken as strange exotic creatures now force their place among 
respectful members of the establishment, as one can see by the number of publications on PBHs per year, which rises 
faster than exponentially. Now PBHs are serious candidates for dark matter carriers, though the original
suggestions~\cite{chaplin} (very light PBH,  $M< 10^{22}$ g) and  \cite{AD-JS} (masses up to millions solar masses, and
even higher) were not taken seriously. 

Presently PBHs not only may solve the problem of dark matter but also nicely solve many other striking problems of modern 
cosmology and astrophysics~\cite{AD-UFN,Carr-conundra}, which are very difficult to solve other way.
 
Discoveries of the last decade indeed demonstrate the contemporary and even the early ($z = 5-10$) universes are indeed unexpectedly 
rich with supermassive black holes (SMBH), ${M= (10^6 - 10^{10}) M_\odot }$
and intermediate mass black holes (IMBH), $M=(10^3 - 10^5)  M_\odot $.

\subsection{Contemporary universe \label{ss-contem}} 

{ Every large galaxy 
contains a central supermassive BH with the
mass  of billions solar masses  in giant elliptical
and compact lenticular galaxies and with a few million solar masses in spiral galaxies, like Milky Way.
{The origin of these BHs is not understood.}
{Accepted faith is that
these BHs are created by matter accretion to a central seed. }
But, the usual accretion efficiency is insufficient by two orders of magnitude to create them during the Universe life-time,
$t_U=14$ Gyr~\cite{mur}. 

Moreover, SMHBs  are discovered  in {tiny galaxies}
 and even in almost {\it empty} space, where is not enough material to create  such huge monsters through accretion,
see review~\cite{AD-UFN} for references.

These observation indicates a possible upside down change of galaxy formation paradigm. Instead of creation of SMBH
by accretion to the galactic center, galaxies are formed by matter attraction by a preexisting primordial black hole as is
suggested in refs.~\cite{AD-JS,DKK,Bosch}.
It is tempting to  make a next step and to conclude that the type of any large galaxy is determined by the mass of  the BH seed.

Another, but possibly far-fetched, piece of evidence in favor of primordial formation of SMBHs is their clumping. There are
at least four binaries of SMBH, one triple system and one quartet. A list of references can be found in  the
review~\cite{AD-UFN}. According to the conventional approach a SMBH binary may be formed in the process of merging
of two galaxies each having a SMBH in the center. It is almost infinitely more problematic to create  a triple system of SMBH
along this way to say nothing about a quartet. It is much easier to create a closely connected system of SMBHs in the early
universe if they are primordial.

Coming to less massive but still huge BHs, note that
about a thousand of black holes with masses $10^5 M_\odot $ and a few with much smaller but still huge masses,
$M\sim 2000 M_\odot$ are discovered presently. It is hard to imagine that such BHs have been created as a result of stellar
collapse. On the other hand the early universe can easily provide them in necessary amount.
According to our prediction~\cite{AD-KP-int-mass},
if the parameters of the mass distribution of PBHs, $M_0$ and $\gamma$, 
are chosen to fit the LIGO data and the normalization factor $\mu$ is chosen to provide
 the density  the PBH with masses exceeding ${10^4 M_\odot}$, which could seed the observed SMBH
in the galaxies, then  the number of PBH with masses 
${(2-3)\times 10^3 M_\odot}$ would be about ${10^4-10^5 }$ per one SMBH. 
This allows all large galaxies to host their own SMBH, seeded by primordial black hole, sometimes even two. 
This predicted density of IMBHs is sufficient to seed the formation of all globular clusters 
and dwarf galaxies. Up to now only one or two IMBHs are observed in globular clusters, while there should be  IMBH in each
of them,
 
\subsection{Anti Dark Matter \label{ss-anti-DM}}

In recent publication, already after the Conference was over, an interesting idea was put forward that dark matter may consist
of compact anti-stars~\cite{anti-DM}. It is noteworthy that  creation of compact stellar-like antimatter objects was 
envisaged in refs.~\cite{AD-JS,DKK}. As argued in refs~\cite{CB-AD,AD-SB,SB-AD-KP}, an abundant density of compact
anti-stars in the universe and even in the Galaxy does not violate existing observational limits. Such anti-DM may be easier to
spot than other forms of macroscopic DM.

\subsection{High $z$ universe \label{ss-young}}

Astronomical observations of the last decade led to the srtiking discovery that 
the early universe at $\bm{z \sim 10}$ is grossly overpopulated with bright quasars or wha tis presumably the same, by SMBHs, 
with masses up to ${M \sim 10^{10} M_\odot}$,} by superluminous young galaxies, by supernovae and gamma-bursters, Moreover, 
the universe at this young age is very dusty and full of heavy elements.

According to ref.~\cite{early-gal}
 the {density of galaxies at ${z \approx 11}$ is 
${10^{-6} }$ Mpc${^{-3}}$, an order of magnitude higher than estimated from the data at lower z.}
It is difficult to understand how these galaxies can be formed using the standard lore.
However, having in our disposal supermassive seeds presented by PBH we can allow for their existence.
So, again first black holes and  then galaxies. 

At present about a hundred quasars at redshift $ z>6$ are known.  The record one at redshift 6.3 has the mass 
equal to twelve billion solar mass~\cite{QSO-12}.
The known accretion rate is not sufficiently strong to ensure their creation. As is shown in 
ref.~\cite{Latif} MBH could accretes only about 2200 ${M_{\odot}}$ during 320 Myr  in the halo 
with a mass of ${ 3 \times 10^{10}~M_{\odot}}$ at $\bm{z=7.5}$. There is general agreement that ultra-massive seeds 
are necessary for the QSO creation, which could be easily presented by PBHs with masses of about $10^4 M_\odot$.

One more argument against sufficiently efficient accretion is neutrality of medium around QSO~\cite{max-z}, 
because otherwise strong accretion should strongly ionize  medium. 

A few months ago a remarkable statement was presented in ref.~\cite{short-time} that quasars at $z\approx 6$ 
remained active only during $10^3 - 10^4$ years. This striking observation implies that we see only  a minor
fraction of SMBH which "lives" in the early universe, not more than  $10^{-4}$. It is hardly possible to create so 
many SMBHs by the conventional accretion mechanism in about 500 million years. PBHs are badly needed.

\section{Conclusion  \label{s-conclud}}

We have have determined  the mass spectrum of black holes using the chirp mass 
distribution extracted from the LIGO data. The best fit spectrum is perfectly well described  
by the log-normal form, confirming the 1993 prediction of ref.~\cite{AD-JS}.

The extracted value of the central mass of the spectrum is equal to 17 solar masses, which very 
well agrees with our earlier result~\cite{AD-KP-mass}. Several astrophysical models of black hole 
creation poorly describe the presented data.

Based on the extracted spectrum parameters one may conclude that practically all black holes in 
the contemporary and early universe are primordial. This presumption allows to solve quite naturally  
a large lot of cosmic conundra.



\section*{Funding}

A.D. acknowledges the support from the Ministry of Science and Higher Education grant\\  No.  FSUS-2020-0039. \\
K.P. acknowledges the support from the RSF Grant 19-42-02004.




\end{document}